\documentclass{article}

\title{Vacuum decay by $p$-branes production\footnote{\textit{To
appear in the Proceedings of the 6th International Symposium
on Frontiers in Fundamental and Computational Physics (FFP6),
September 26-29, Udine, ITALY}}}

\author{Lorenzo Sindoni\footnote{Email: \texttt{potenzo17@yahoo.it}}\\
{\small{}Dipartimento di Fisica,}
{\small{}Universit\`{a} degli Studi di Trieste,}\\
{\small{}via A. Valerio, 2 - I-34127 Trieste (TS), ITALY}\\[3mm]
and\\[3mm]
Stefano Ansoldi\footnote{Email: \texttt{ansoldi@trieste.infn.it}}%
{~}\footnote{Webpage: \texttt{http://www-dft.ts.infn.it/$\sim$ansoldi}}\\
{\small{}Dipartimento di Matematica e Informatica,}
{\small{}Universit\`{a} degli Studi di Udine,}\\
{\small{}and I.N.F.N. - Sezione di Trieste}\\
{\small{}via delle Scienze, 206 - I-33100 Udine (UD), ITALY}}

\begin{document}

\maketitle

\begin{abstract}
We present a generalization to the $N$-dimensional case for
the nucleation coefficient of a spherical $p$-brane,
separating two (anti-)de Sitter spacetimes.
We use a semiclassical approximation based on the analytical
continuation to the Euclidean sector of a suitable effective
action describing a $p$-brane in General Relativity.
\end{abstract}

\section{Introduction}

Vacuum decay can be seen as a phase transition in spacetime
and a long time ago the relevance of gravity for the process
was studied \cite{bib:ColDeL}. The standard treatment
of this process makes use of a scalar field, known as the
{\em inflaton}, that drives the transition between the false and true
vacuum states. This situation can be described by instanton
calculations as, for instance, the Coleman-de Luccia and the
Hawking-Moss instantons.

Here we present a different approach, generalizing past works of
one of the authors \cite{reviews,cqsdvd}. In particular we are going to use
(anti-)de Sitter solutions in $N$ spacetime dimensions.
In this background we put a spherically symmetric $(N-1)$-brane
that splits spacetime into two domains. The system can be described
by Israel junction conditions \cite{Israel}, which provide the equations of motion
for the timelike brane. The associated solutions are of two kinds:
the first one consists of a degenerate brane of zero radius,
while the second one consists of a \emph{bounce} brane collapsing from infinity
towards a finite nonzero turning point, and then re-expanding.
To model vacuum decay we consider the tunnelling from the
zero radius solution to the bounce solution. The corresponding physical
picture is the following: a very small brane\footnote{In the mathematical
treatment of the classical situation the brane has, in fact, zero radius; from
the physical point of view, with quantum gravity in mind, we can imagine this
brane as a result of zero point quantum fluctuations.} inside a de
Sitter geometry with cosmological constant $\Lambda _{+}$, due to quantum effects,
has a non-vanishing probability to \emph{tunnel} into a brane, containing
a de Sitter spacetime with a different cosmological constant $\Lambda _{-}$.
This represents the formation of a bubble of a different vacuum phase that then
expands to infinity, realizing a transition of the whole spacetime geometry.
We can obtain an expression for the probability of such
a process using an effective action for this system.

\section{Classical Dynamics}

The stress-energy tensor for a distribution of matter localized on
an hypersurface $\Sigma$ (the $p$-brane we mentioned above)
can be written in the form $S_{\mu\nu} \delta(\eta)$,
where $\delta$ is a Dirac delta, and $\eta$ can be thought as
a transverse coordinate to $\Sigma$.
In $N$-dimensional General Relativity it is possible to write down
the equations of motion for this infinitesimally thin distribution of
matter by splitting Einstein equations in the tangential and
transverse part (see \cite{Israel} for the $4$-dimensional case; it
can be extended to higher dimensions). Israel junction conditions, then, are
\[
\left[ {\cal K}_{ij} - h_{ij} {\cal K} \right] \propto S_{ij}
,
\]
where ${\cal K}_{ij}$ and $\cal K$ are, respectively, the extrinsic
curvature tensor and its trace and $h_{ij}$ is the induced metric
on $\Sigma$. Here we introduced the standard notation
$[A] = \lim_{\eta\rightarrow0^+} \{A(\eta) - A(-\eta)\}$.
Israel junction conditions describe how the
$(N-1)$-brane is embedded in the (in principle different) geometries of the
two spacetime domains that it separates.
For our purposes, we are going to write down these equations for a spherical
brane with surface stress energy tensor $S _{ij} = k h _{ij}$
separating two de Sitter spacetimes.
This can be done explicitly
in terms of the radius $R$ of the brane\footnote{We are going to consider $R$
as a function $R (\tau)$ of the proper time $\tau$ of an observer sitting on the
brane and denote by an overdot the derivative with respect to $\tau$.}.
Then Israel junction conditions reduce to the single differential equation
\begin{equation}
{\cal H}(R,\dot{R})=\left[ \epsilon \sqrt{\dot{R}^2 + f(R)}
\right]R^{(N-3)} - kR^{N-2} = 0
;
\label{eq:sphjuncon}
\end{equation}
$k$ is the constant \textit{tension} of the brane, $\epsilon$ are
signs to be determined by the equation itself \cite{tesi}, and
$f(R)= 1 - \Lambda R^2$
is the metric function appearing in the static line element adapted to the
spherical symmetry for the (anti-)de Sitter spacetime\footnote{Since the compact
notation could be misleading, we remember that we have two spacetimes with different
cosmological constants $\Lambda _{\pm}$ and, thus, two metric functions $f _{\pm} (R)$;
please, also remember the meaning of the square brackets defined above.}.
For suitable values of the cosmological constants
equation (\ref{eq:sphjuncon}) has two types of solutions: the first is $R \equiv 0$,
while the second represents a brane collapsing and re-expanding from and
to infinity. For our purpose it is also important to note that
equation (\ref{eq:sphjuncon}) can also be obtained by an effective action,
which in the $N$-dimensional case can be written as
\begin{equation}
S_{\mathrm{eff.}} =\int \left\{ R^{N-3} \dot{R} \left[ \chi {\rm th}^{-1} \left(
\frac{\dot{R}}{\sqrt{\dot{R}^2+f(R)}} \right) \right] - {\cal H}(R,\dot{R})
\right \}
d\tau
,
\label{eq:actfun}
\end{equation}
with the additional {\em constraint} ${\cal H} = 0$ that has to be imposed on the solutions
of the corresponding Euler-Lagrange equation \cite{tesi}.

\section{Tunnelling}

The action (\ref{eq:actfun}) is crucial in our semiclassical quantization
program, since it can be used to quantize the system \emph{via} a path integral
approach. Here we are going to consider the tunnelling
process from the $R \equiv 0$ solution to the bounce solution, within the
saddle-point approximation. This gives the possibility to estimate the
following approximated amplitude
\begin{equation}
A_{\mathrm{s.p.}} \propto \exp \left( - S^{(\mathrm{e})}_{\mathrm{eff.}} \right)
,
\label{eq:eucact}
\end{equation}
where $S^{(\mathrm{e})}_{\mathrm{eff.}}$ is the Euclidean effective action obtained by
analytically continuing the action (\ref{eq:actfun}) to the
Euclidean sector. In order to simplify some expressions, we introduce
the following adimensional quantities:
\begin{displaymath}
x = k R
\quad , \quad
t = k \tau
\quad , \quad
\alpha = \frac{\Lambda_{-} + \Lambda_{+}}{k^2}
\quad , \quad
\beta = \frac{\Lambda_{-} - \Lambda_{+}}{k^2}.
\end{displaymath}
Moreover it is a well known result that the adimensional version of the
equation of motion (\ref{eq:sphjuncon}) can be cast in the following form
\[
(x ')^2 + V(x) = 0
,
\]
where the prime now denotes the derivative with respect the adimensional time $t$.
The potential $V (x)$ is given by
\begin{equation}
V(x) = 1 - \frac{x^2}{x _{0} ^{2}}
,
\label{eq:effpot}
\end{equation}
where $x_0=2/\sqrt{(1+2\alpha+\beta^2)}$ is the adimensional turning radius,
provided that the argument of the square root is positive.
If this condition holds, there is a bounce trajectory, otherwise we have only
the $x=0$ solution. Starting from (\ref{eq:actfun}), (\ref{eq:eucact}) and
(\ref{eq:effpot}) it is possible to evaluate the Euclidean action on the
tunnelling trajectory, i.e on the segment $[0,x _{0}]$. The final result can
be expressed as
\begin{equation}
S^{(\mathrm{e})}_{\mathrm{eff}}=\frac{x_0 ^{(N-3)/2}}{2(N-2)k^{(N-2)}} \left[
(\beta-\epsilon) J(N,C_{\sigma}) \right],
\end{equation}
where $J (N,p)$ is a function of the dimension of spacetime, $N$, and of
\[
C_{\sigma} = 1 - \left(\frac{\beta - \sigma}{2}\right)^2 x_0^2; \\
\]
with
\[
    \sigma _{\pm} = \pm 1
    .
\]
A detailed description of the functional form of $J (N , p)$ is beyond the
scope of this contribution and can be found elsewhere \cite{bib:SinAnsToPub}.
We just remark one important feature of it, namely that $J (N , p)$ is defined
only when $p < 1$, a condition that, according to the form of $C _{\sigma}$,
is always satisfied in our case.

\section{Conclusions}

In this contribution we have summarized how it is possible to compute
the tunnelling amplitude for a spherical $p$-brane: the process, in view of
our short discussion in the introduction, can be used to model the transition
between two different vacuum phases, one being the de Sitter
spacetime with cosmological constant $\Lambda _{+}$ and the other
being the inflating inside of the $p$-brane, which is characterized
by the cosmological constant $\Lambda _{-}$. This computation was already
performed in four dimensions in the seminal work by Coleman and de Luccia
\cite{bib:ColDeL} and that result was reproduced in \cite{cqsdvd}. Here
we have carried out its generalization to arbitrary dimensions (referring to
\cite{bib:SinAnsToPub} for a more detailed analysis).

\paragraph{ACKNOWLEDGMENTS} One of us (L.S.) wants to thank the
Physics Department of University of Udine for financial support.

\end{document}